\documentstyle[12pt]{article}
\begin{document}
\renewcommand{\baselinestretch}{1.5}

\renewcommand{\baselinestretch}{1.5}
\newcommand\beq{\begin{equation}}
\newcommand\eeq{\end{equation}}
\newcommand\bea{\begin{eqnarray}}
\newcommand\eea{\end{eqnarray}}

\centerline{\bf SEMICLASSICAL AND FIELD THEORETIC STUDIES OF}
\centerline{\bf HEISENBERG ANTIFERROMAGNETIC CHAINS WITH}
\centerline{\bf FRUSTRATION AND DIMERIZATION}

\vskip 1 true cm

\centerline{Sumathi Rao \footnote{{\it E-mail address}: 
sumathi@mri.ernet.in}} 
\centerline{\it Mehta Research Institute, 10 Kasturba Gandhi Marg,}
\centerline{\it Allahabad 211002, India}
\vskip .5 true cm

\centerline{Diptiman Sen \footnote{{\it E-mail address}:
diptiman@cts.iisc.ernet.in}} 
\centerline{\it Centre for Theoretical Studies, Indian Institute
of Science,}
\centerline{\it Bangalore 560012, India}
\vskip 2 true cm

\noindent {\bf Abstract}
\vskip 1 true cm

The Heisenberg antiferromagnetic spin chain with both
dimerization and frustration is studied. The classical ground state has
three phases (a Neel phase, a spiral phase and a colinear phase), around
which a planar spin-wave analysis is performed. In each phase, we discuss a 
non-linear sigma model field theory describing the low energy excitations. A
renormalization group analysis of the $SO(3)$ matrix-valued field theory 
of the spiral phase leads to the conclusion that the theory becomes $SO(3) 
\times SO(3)$ and Lorentz invariant at long distances. This theory is 
analytically known to have a
massive spin-$1/2$ excitation. We also show that $Z_2 ~$ solitons in the field 
theory  lead to a double degeneracy in the spectrum for half-integer spins.

\vskip 1 true cm

\noindent PACS numbers: ~75.10.Jm, ~75.50.Ee, ~11.10.Lm

\newpage

\leftline{\bf I. INTRODUCTION}
\vskip .5 true cm

Antiferromagnets in low dimensions have continued to draw considerable interest 
in recent years, mainly because of their possible relevance to
high $T_c$ superconductors. A large  variety of
theoretical tools have become available to study them  including
non-linear sigma model (NLSM) field theories [1-6], Schwinger boson 
\cite{SB} and other bosonic mean field theories \cite{OTHERB}, fermionic mean
field theories \cite{AM}, series expansions \cite{SERIES}, exact
diagonalization of small systems \cite{EXACT}, and the density
matrix renormalization group (DMRG) method \cite{DMRG,CHITRA}. In
particular, one of the most useful tools has been the use of 
the non-linear sigma model (NLSM) field theories [1-6].
These theories display interesting features such as
dynamical mass generation and existence of topological soliton solutions,
which are of relevance in spin models. In fact, it was the mapping of the
$O(3)$ NLSM to the Heisenberg antiferromagnet which led Haldane
\cite{HALDANE} to his celebrated conjecture
that integer spin models would have a gap, contrary to the known
solution for the spin-$1/2$ model. The experimental verification 
\cite{HGAPEXPT} of this conjecture has given special impetus to the study of 
NLSM field theories for spin models.
Ever since then, various generalizations of the Heisenberg
antiferromagnetic spin chain including models with dimerization 
and/or frustration \cite{RAO2} and coupled chains or Heisenberg
ladders \cite{CCLAD} have been studied.

In this paper, we study a general Heisenberg spin chain with
both dimerization (an alternation $\delta$ of the nearest
neighbor (nn) couplings) and frustration (a next-nearest
neighbor (nnn) coupling $J_2 ~$). Even in the classical 
spin $S \rightarrow \infty$ limit, the system has
a rich ground state `phase diagram', \cite{FN1} with three distinct phases,
a Neel phase, a spiral phase and a colinear phase (defined
below). For large, but finite $S$, a spin-wave analysis {\it a la} Villain 
\cite{VILLAIN} can be performed which gives the spectrum of low energy 
excitations about the classical ground state. This is described in Sec. II.

The Neel phase via the Haldane mapping to an $O(3)$ NLSM is well-known to be 
gapless for half-integer spins but gapped for integer spins 
\cite{HALDANE,AFFLECK}. An
interesting and important question to address is whether this distinction
persists in the spiral and colinear phases as well.
To answer this, we need to describe the long wavelength
fluctuations about the classical ground state in these phases as well
by non-linear field theories which are capable of systematic improvement
by renormalization group techniques. This has been identified in
the spiral phase (for $\delta = 0$) \cite{RAO1,ALLEN}, and 
in the colinear phase for $\delta =1$, which
corresponds to two coupled spin chains \cite{RAO2,CCLAD}. However, while the
Neel phase has been extensively studied, various aspects like
the ground state degeneracy and the low energy spectrum are not
well understood in the spiral and colinear phases. 

For completeness and to set the notation, we first review in Sec. III A, 
the NLSM and the RG in the Neel phase for
arbitrary $J_2$ and $\delta$. For the spiral phase, in Sec. III B, we obtain 
the $SO(3)$-valued non-linear field theory and show using a (previously
derived) one-loop $\beta$-function that the
field theory flows to an $SO(3) \times SO(3)$ symmetric and
Lorentz invariant theory with an analytically known spectrum
\cite{ANALYTIC}.  However, this does not directly lead to a distinction between
integer and half-integer spins. But in Sec. III C, we discuss how the 
presence of $Z_2 ~$ solitons affects the ground state
degeneracy and the low energy spectrum and 
does provide a distinction. We are led to the conclusion that the system
is gapped both for integer and non-integer spins, but the ground state is 
degenerate for half-integer spins (when there is no dimerization).
In Sec. III D, we briefly discuss the NLSM for a spin ladder. It is
qualitatively similar to the one in the Neel phase. However, no topological
term is induced and there is no distinction between
the integer and half-integer spin cases. A shorter 
version of these results will appear elsewhere \cite{RAO2}.
Finally, we make some concluding remarks in Sec. IV. 

\vskip .5 true cm
\leftline{\bf II. CLASSICAL PHASE DIAGRAM AND SPIN WAVE} 
\leftline{\bf ~~~~ ANALYSIS} 
\vskip .5 true cm

The Hamiltonian for the frustrated and dimerized spin chain is given by 
\begin{equation}
H ~=~ J_1 ~\sum_i ~(~1~+~(-1)^i \delta ~) {\bf S}_i \cdot {\bf
S}_{i+1} ~ +~ J_2 ~ \sum_i ~{\bf S}_i \cdot {\bf S}_{i+2} ~,
\label{ham1}
\end{equation}
where ${\bf S}_i^2 = S(S+1) \hbar^2$, the coupling constants
$J_1, J_2 \ge 0$ and the dimerization parameter $\delta$ lies
between $0$ and $1$. (We will henceforth set $\hbar =1$). 
Classically (for $S \rightarrow \infty$), the ground state is a coplanar 
configuration of the spins with energy per spin equal to
\begin{equation}
E_0 ~=~ S^2 ~\left[ {J_1 \over 2} ~ (1+\delta) \cos \theta_1 ~+~
{J_1 \over 2} ~ (1-\delta ) \cos \theta_2 ~+~ J_2 \cos (\theta_1
+ \theta_2) \right]~, 
\end{equation}
where $\theta_1$ is the angle between the spins ${\bf S}_{2i}$
and ${\bf S}_{2i+1}$ and $\theta_2$ is the angle between the
spins ${\bf S}_{2i}$ and ${\bf S}_{2i-1}$.  Minimization of the
classical energy with respect to the $\theta_i ~$ yields the following phases.

\noindent (i) Neel: This phase has $\theta_1 = \theta_2 = \pi$
and is stable for $1-\delta^2 > 4J_2/J_1$.

\noindent (ii) Spiral: Here, the angles $\theta_1$ and $\theta_2$ are given by
\begin{eqnarray}
\cos \theta_1 &=& - {1\over{1+\delta}}~ {\left[~ {{1 - \delta^2}
\over{4 J_2/J_1}} ~+~ {\delta \over {1 + \delta^2}}~ {4 J_2
\over J_1} ~ \right]} \nonumber \\ 
{\rm and} \quad \cos \theta_2 &=& - {1\over{1 -\delta}}~ {\left[~ {{1 -
\delta^2} \over {4 J_2/J_1}} ~-~ {\delta \over {1 - \delta^2}}~
{4 J_2 \over J_1}  ~ \right]},
\end{eqnarray}
where $\pi/2 < \theta_1 <\pi$ and $0<\theta_2 < \theta_1$. This phase
is stable for $1-\delta^2 < 4 J_2/J_1 < (1-\delta^2) /\delta$.

\noindent (iii) Colinear: This phase (which needs both
dimerization and frustration) is defined to have $\theta_1 =
\pi$ and $\theta_2 = 0$. It is stable for $(1-\delta^2) /\delta < 4 J_2/J_1$. 

\noindent These phases along with the phase boundaries are depicted in Fig. 1.

We now study the spin wave spectrum about the ground state.
Before describing the details of the calculations, we state the main results.
In
the Neel phase, we find two zero modes with equal velocities. In
the spiral phase, we have three modes, two with the same
velocity describing out-of-plane fluctuations and one with a
higher velocity describing in-plane fluctuations.  In the
colinear phase, we get two zero modes with equal velocities just
as in the Neel phase.  The three phases also differ
in the behavior of the spin-spin correlation function 
$S(q) = \sum_n \langle {\vec S}_o \cdot {\vec S}_n \rangle \exp (-iqn)$ 
in the classical limit. $S(q)$ is peaked at $q = (\theta_1 + \theta_2 )/2$, 
{\it i.e.}, at $q=\pi$ in the Neel phase, at $\pi/2 < q < \pi$ in the spiral
phase and at $q=\pi/2$ in the colinear phase.  Even for $S=1/2$
and $1$, DMRG studies have seen this feature of $S(q)$ in the Neel and spiral 
phases \cite{CHITRA}. The colinear region has not yet been probed numerically. 

Since the classical ground state is coplanar (with the spins lying in, say, 
the ${\hat {\bf x}}-{\hat {\bf y}}$ plane) in all the phases, it is more 
convenient
to develop the spin wave theory by parametrizing the spin operators in terms
of Villain's variables \cite{VILLAIN}, rather than the usual Holstein-Primakoff
variables \cite{HP}. The spin variables at each site are expressed in terms of
two conjugate variables $S_{iz}$ and $\phi_i$ as
\bea
S_i^+ ~&=&~ \exp ~(i \phi_i ) ~\Bigl[ ~\Bigl(S+ {1 \over 2} \Bigr)^2 ~-~ 
\Bigl(S_{iz} + {1 \over 2} \Bigr)^2 ~\Bigr]^{1/2} ~, \nonumber \\
S_i^- ~&=&~ ~\Bigl[ ~\Bigl(S+ {1 \over 2} \Bigr)^2 ~-~ \Bigl(S_{iz} + 
{1 \over 2} \Bigr)^2 ~ \Bigr]^{1/2} \exp ~(- i \phi_i ) ~, \nonumber \\
{\rm and} \quad S_i^z ~&=&~ {1 \over i} ~{{\partial} \over {\partial \phi_i}} ~.
\label{villain}
\eea
The periodic operator $\phi_i$ (with $\phi_i \equiv \phi_i + 2 \pi$) satisfies
the commutation relation
\beq
\Bigl[ ~\phi_i ~,~{{S_{jz}} \over S} ~\Bigr] ~=~ {i \over S} ~\delta_{ij} ~.
\label{comm1}
\eeq
In the large-$S$ limit, the discreteness of $S_{iz} / S$ and the periodicity
of $\phi_i$ can be ignored. $S_{iz}$ and $\phi_i$ can then be thought of as
canonically conjugate and continuous momentum and position operators. In the
classical limit $S \rightarrow \infty$, $S_{iz} /S \rightarrow 0$ and $\phi_i$
is fixed at a value ${\overline \phi}_i$ which is the angle made by the
classical spin ${\bf S}_i$ relative to the $\bf x$-axis in the plane. 

We now expand the square roots in Eq. (\ref{villain}) and keep terms only up 
to order $1/S$. On expanding the angles $\phi_i$ to quadratic order about 
${\overline \phi}_i$, we obtain
\bea
S_i^+ ~&=&~ \exp ~(i {\overline \phi}_i ) ~(S + {1 \over 2}) ~(1 + i \sigma_i
- {1 \over 2} \sigma_i^2 ) ~\Bigl( 1 - {1 \over 2} {{(S_{iz} + {1 \over 2})^2}
\over {(S + {1 \over 2})^2}} ~\Bigr) ~, \nonumber \\
S_i^- ~&=&~ \exp ~(- i {\overline \phi}_i ) ~(S + {1 \over 2}) ~(1 - i \sigma_i
- {1 \over 2} \sigma_i^2 ) ~\Bigl( 1 - {1 \over 2} {{(S_{iz} + {1 \over 2})^2}
\over {(S + {1 \over 2})^2}} ~\Bigr) ~, 
\label{expand}
\eea
where $\sigma_i$ describes in-plane fluctuations (about the angle ${\overline 
\phi}_i$) and $S_{iz} /S$ describes out-of-plane fluctuations. In the
presence of dimerization $\delta$, the unit cell contains $2$ sites and has
size $2a$, where $a$ is the lattice spacing. We therefore define
\bea
S_n^{(1)} ~&=&~ S_{2n+1,z} ~, \quad S_n^{(2)} ~=~ S_{2n,z} ~, \nonumber \\
{\rm and} \quad \sigma_n^{(1)} ~&=&~ \sigma_{2n+1} ~, \quad 
\sigma_n^{(2)} ~=~ \sigma_{2n} ~. 
\label{def}
\eea
and their Fourier components
\bea
S_k^{(a)} ~&=&~ {\sqrt {2 \over N}} ~\sum_n ~S_n^{(a)} \exp (-i2k na) ~,
\nonumber \\
{\rm and} \quad \sigma_k^{(a)} ~&=&~ {\sqrt {2 \over N}} ~\sum_n ~
\sigma_n^{(a)} \exp (-i2k na) ~,
\label{fourier}
\eea
where $a=1,2$, and 
\beq
[ ~\sigma_k^{(a)} ~,~ S_{k^\prime}^{(b)} ~] ~=~ i \delta^{ab} \delta_{k+
k^{\prime},0} ~
\label{comm2}
\eeq
for the doubled unit cell. Substituting (\ref{expand}) in terms of the Fourier 
components of the doubled unit cell in (\ref{ham1}), we find
\beq
H ~=~ E_o ~+~ \sum_{k} \sum_{a,b} ~\Bigl[ ~S_k^{(a)} 
{\underline A}_k^{ab} S_{-k}^{(b)} ~+~ S^2 ~\sigma_k^{(a)} 
{\underline B}_k^{ab} \sigma_{-k}^{(b)} ~\Bigr] ~,
\label{ham2}
\eeq
where $0<k< \pi /2a$ and $a,b=1,2$; and ${\underline A}_k$ 
and ${\underline B}_k$ are $2 \times 2$ hermitian matrices of the form
\bea
{\underline A}_k ~=~ & & \left({\begin{array}{cc}
                                 a_1 & a_2 + a_3 e^{i2ka} \\
                                 a_2 + a_3 e^{-i2ka} & a_1 \\
                                \end{array}} \right) ~, \nonumber \\
{\rm with} \quad a_1 ~=~ &-& ~J_1 ~[~ ( 1 + \delta ) \cos \theta_1 ~+~ ( 1 -
\delta ) \cos \theta_2 ~] \nonumber \\
&-& ~2 J_2 \cos ( \theta_1 + \theta_2 ) ~+~ 2 J_2 \cos 2ka ~, \nonumber \\
a_2 ~=~ &J_1& ( 1 + \delta ) ~, \quad a_3 ~=~ J_1 ( 1 - \delta ) ~, 
\label{amatrix}
\eea
and
\bea
{\underline B}_k ~=~ & & \left({\begin{array}{cc}
                                 b_1 & b_2 + b_3 e^{i2ka} \\
                                 b_2 + b_3 e^{-i2ka} & b_1 \\
                                \end{array}} \right) ~, \nonumber \\
{\rm with} \quad b_1 ~=~ &-& ~J_1 ~[~ ( 1 + \delta ) \cos \theta_1 ~+~ ( 1 -
\delta ) \cos \theta_2 ~] \nonumber \\
&-& ~2 J_2 \cos ( \theta_1 + \theta_2 ) ~+~ 2 J_2 
\cos (\theta_1 + \theta_2 ) ~ \cos 2ka ~, \nonumber \\
b_2 ~=~ &J_1& ( 1 + \delta ) \cos \theta_1 ~, \quad b_3 ~=~ J_1 ( 1 - 
\delta ) \cos \theta_2 ~. 
\label{bmatrix}
\eea
These formulae hold for all three phases; the angles $\theta_i$ were given 
earlier for each phase. To find the spin wave spectrum $\omega (k)$, we have to 
diagonalize ${\underline A}_k$ and ${\underline B}_k$, {\it i.e.}, we have to 
find a matrix $U$ (not necessarily unitary) such that ${\rm det}~~ U = 1$,
and $U A_k U^{\dag}$ and $U^{-1 \dag} B_k U^{-1}$ are real diagonal matrices
with entries diag$~~(\alpha_1 , \alpha_2)$ and 
diag$~~(\beta_1 , \beta_2)$ respectively.
Then the energies of modes of momentum $k$ are given by $\omega_{1k} =
S {\sqrt {\alpha_1 \beta_1}}$ and $\omega_{2k} = S {\sqrt {\alpha_2 \beta_2}}$.
(There are two modes for each $k$ because a unit cell contains two sites).

In the Neel phase, both the energies vanish at $k=0$. The spin wave velocity $c$
is defined to be the derivative $\partial \omega / \partial k$ at the gapless
point. We find that $c = 2 J_1 Sa {\sqrt {1 - \delta^2 - 4J_2 /J_1}}$ for
both branches. In the spiral phase, we find one gapless mode at $k=0$ 
(with a slope $c_o$) and two gapless modes at $2 ka = 2 \pi - \theta_1 - 
\theta_2$. The latter are counted as
two modes because the energy vanishes as one approaches that value of $k$ both
from the left and from the right; and the two slopes $c_1$ are equal. 
The expressions for these velocities are quite complicated for general 
$\delta$. (However, we find that the
spin wave velocity $c_o$ is always greater than $c_1$). For $\delta =0$, 
the expressions simplify to give $c_o = J_1 S a (1+4 J_2 /J_1) {\sqrt {1 -
J_1^2 /16 J_2^2}}$ and $c_1 = c_o {\sqrt {[1+(1 - J_1 /2 J_2)^2 ]/2}}$.
Finally, in the colinear phase, one of the energies vanishes at $k=0$ and
the other at $2ka = \pi$. The spin wave velocities are equal at the two
points and are given by $c= 2 J_1 S a {\sqrt {(4J_2 \delta /J_1 + \delta^2 -1)
(2J_2 /J_1 + \delta + 1)/(2 \delta)}}$. One can also check that in 
all the cases, the spin wave velocities vanish as 
we approach the boundary separating any two phases.

The above calculation of the spin wave velocities $c$ assumes long range order. 
For finite values of $S$, the spin chain models actually have no long range 
order, {\it i.e.}, the two-spin correlation function goes to zero at 
large separation either algebraically or exponentially. However, it is still
useful to calculate $c$ because it is one of the parameters which enters
the NLSM field theories; one therefore has a consistency check on the 
derivation of the field theories.

\vskip .5 true cm
\leftline{\bf III. NLSM FIELD THEORIES}
\vskip .5 true cm
\leftline{\bf A. Neel Phase}
\vskip .5 true cm

To study non-perturbative aspects, the NLSM approach is convenient since the
RG can be used to improve naive perturbation results. The NLSM is well-known 
in the Neel phase \cite{HALDANE,AFFLECK} and can be most easily derived as 
follows. The field variable is a unit vector ${\vec \phi}$. Since the classical
ground state has a periodicity of two sites, we group the spins in two's and
define
\bea
{\vec \ell}_{2i+1/2} ~&=&~ {{{\bf S}_{2i} + {\bf S}_{2i+1}} \over {2a}} ~,
\nonumber \\
{\rm and} \quad {\vec \phi}_{2i+1/2} ~&=&~ {{{\bf S}_{2i} - {\bf S}_{2i+1}} 
\over {2S}} ~.
\label{phi1}
\eea
These variables satisfy the identites
\bea
{\vec \ell} \cdot {\vec \phi} ~&=&~ 0 ~, \nonumber \\
{\rm and} \quad {\vec \phi}^2 ~&=&~ 1 ~+~ {1 \over S} ~-~ {{a^2 {\vec \ell}^2} 
\over {S^2}} ~,
\label{constraint}
\eea
so that ${\vec \phi}^2 = 1$ in the large-$S$ and long wavelength limit. (We 
will see below that $\vec \ell$ is proportional to a single space-time 
derivative of $\vec \phi$; hence long wavelength means that $\vert a {\vec 
\ell} \vert << 1$). We 
define the continuum space coordinate $x=2ia$ so that $\delta_{2i,2j} /2a = 
\delta (x-y)$. The variables in Eq. (\ref{phi1}) therefore satisfy the 
commutation relations
\bea
\left[~ \ell_{\alpha} (x) ~,~ \ell_{\beta} (y) ~\right] ~&=&~ i ~
\epsilon_{\alpha \beta \gamma}~ \ell_{\gamma} (x)~ \delta 
(x-y) ~, \nonumber \\ 
\left[~ \ell_{\alpha} (x) ~,~ \phi_{\beta} (y) ~\right] ~&=&~ i ~
\epsilon_{\alpha \beta \gamma}~ \phi_{\gamma} (x)~ \delta 
(x-y) ~, \nonumber \\ 
{\rm and} \quad [~\phi_{\alpha} (x) ~,~ \phi_{\beta} (y) ~] ~&=&~ 0 ~.
\label{comm3}
\eea
where $\alpha , \beta , \gamma = x,y,z$, and $\gamma$ is summed over on the 
right hand sides. Thus ${\vec \ell} (x)$ can be identified as the angular 
momentum of the field $\vec \phi$, and ${\vec \ell} = {\vec \phi} \times {\vec 
\pi}$, where $\vec \pi$ is the momentum canonically conjugate to $\vec \phi$. 
The Hamiltonian (\ref{ham1}) can be rewritten in terms of $\vec \ell$ and 
$\vec \phi$, and Taylor expanded in space derivatives, {\it e.g.},
\beq
{\vec \phi}_{2i+5/2} ~=~ {\vec \phi}_{2i+1/2} ~+~ 2a ~{\vec \phi}_{2i+1/2}^
{\prime} ~+~ 2a^2 ~{\vec \phi}_{2i+1/2}^{\prime \prime} ~+~ ... ~.
\label{expansion}
\eeq
For $\vec \ell$, since it is already proportional to $\dot {\vec\phi}$, 
we only need to keep terms up to ${\vec \ell}^2$, 
$\dot {\vec \ell}$ and ${\vec \ell}^{\prime}$,
where the dot and prime denote time and space derivatives respectively. Finally,
we replace the sum $\sum_{2i} = \int dx /2a$ to obtain a continuum Hamiltonian
\beq
H ~=~ \int ~dx ~\Bigl[ ~{{c g^2} \over 2} ~\Bigl({\vec \ell} + {S \over 2} 
(1 - \delta) {\vec \phi}^{\prime} \Bigr)^2 ~+~ {c \over {2g^2}} ~
{\vec \phi}^{\prime 2} ~\Bigr] ~,
\label{ham3}
\eeq
where $c$ is the spin wave velocity obtained earlier in the Neel
phase, and the coupling constant $g^2  = 2/(S \sqrt{1-\delta^2 - 4 J_2/J_1} )$.
Thus large-$S$ corresponds to weak coupling. 

The Hamiltonian in (\ref{ham3}) follows from the `Lorentz-invariant' 
Lagrangian density 
\beq
{\cal L} ~=~ {{\dot {\vec \phi}}^2 \over {2 c g^2}} ~-~ {{c {\vec \phi}^{\prime 
2}} \over {2g^2}} ~+~ {\theta \over 4 \pi}
{}~{\vec \phi} \cdot {\dot {\vec \phi}} \times {\vec \phi}^{\prime} ~. 
\label{lag1}
\eeq
Here $c = 2 J_1 aS \sqrt{1-\delta^2 -4 J_2/J_1}~$ is the spin wave velocity 
($a$ is the lattice spacing) and $g^2$ is the coupling constant. 
The third term in (4) is a topological term with $\theta = 2 \pi S (1-\delta)$. 
It is known that this field theory is gapless for $\theta = \pi$ mod $2\pi$ 
with the correlation function falling off with the power $1$ at large 
separations, and is gapped otherwise \cite{AFFLECK}. 
For the gapped theory, the correlations decay exponentially with correlation
length $\zeta$, where $\zeta$ is found from a one-loop RG calculation to
be $\zeta /a = \exp (2\pi/g^2)$. Hence $\ln (\zeta/a) = \pi S \sqrt{1 -
\delta^2 - 4 J_2/J_1}$. For completeness, this is plotted in Fig. 1 for 
$\delta =0$ and $4 J_2/J_1<1$. 

\newpage
\leftline{\bf B. Spiral Phase}
\vskip .5 true cm

Recently, the NLSM for the spiral phase has been studied for $\delta =0$
\cite{RAO1,ALLEN}. Since the classical ground state is generally not periodic,
we will follow the treatment of Ref. \cite {ALLEN}. The ground state has 
$\theta_1 = \theta_2 
= \theta = \cos^{-1} (- J_1/4 J_2)$. The field variable describing
fluctuations about the classical ground state is an $SO(3)$
matrix ${\underline R} (x,t)$ related to the spin variable at
the $i^{\rm th}$ site as $({\bf S}_i)_a = S {\underline
R}_{ab} {\hat {\bf n}}_b ~$, where $a,b = 1,2,3$ are the components
along the ${\hat {\bf x}}$, ${\hat {\bf y}}$ and ${\hat {\bf
z}}$ axis, and ${\hat {\bf n}}$ is a unit vector given by
\beq
{\hat {\bf n}}_i ~=~ {{{\hat {\bf x}} \cos i\theta + {\hat {\bf y}}
\sin i \theta + a{{\vec \ell}}} \over {\vert ~{\hat {\bf x}}
\cos i\theta + {\hat {\bf y}} \sin i \theta + a{{\vec \ell}}
~\vert}} ~. 
\label{nvector}
\eeq
The unit vector ${\hat {\bf n}}_i ~$ describes the orientation of the $i^{\rm 
th}$ spin in the classical ground state, and ${a \vec \ell}$ represents the 
small deviation from the classical configuration (it is {\it not}
the angular momentum as in our discussion of the Neel phase in 
(\ref{phi1}) ~). The Hamiltonian in Eq. (\ref{ham1}) 
can be expanded in terms of ${\underline R}$ and ${{\vec \ell}}$, and 
Taylor expanded up to second order in space-time derivatives to obtain a 
continuum Hamiltonian \cite{ALLEN}
\beq
H ~=~ \sum_{a,b} ~\ell_a M_{ab} \ell_b ~+~ {\rm tr} ~( 
{\underline R}^{\prime T} {\underline R}^{\prime} P) ~.
\label{ham4}
\eeq
Here $M$ and $P$ are diagonal matrices with 
\bea
M & = & {\rm diag} ~~( M_{11}, M_{22}, M_{33} ) \nonumber\\
{\rm with} \quad M_{33} &=& 2 J_2 a S^2 (1+J_1/4J_2)^2 , \nonumber\\
{\rm and}\quad  M_{11}=M_{22}&=&M_{33} [1+(1-J_1/2J_2)^2]/4,  
\eea
and 
\bea
P &=&{\rm diag}~~ (P_{11}, P_{22}, P_{33} ) \nonumber\\
{\rm with} \quad P_{11} =
P_{22} &=& J_2 a S^2 (1-J_1^2/16J_2^2),\nonumber\\
{\rm and} \quad P_{33}&=&0. 
\eea
The Lagrangian is then 
\beq
{\cal L}={\cal L}_K - H,  \quad
{\rm with} \quad {\cal L}_K = a S ({\underline T} 
{\vec \ell}) \cdot {\bf V} 
\eeq 
where 
\beq
{\underline T}={\rm diag} ~~
(1/2,1/2,1) \quad {\rm and} \quad V_a = -1/2 
\epsilon_{abc} ({\underline R}^T {\dot {\underline R}})_{bc}.
\eeq 
${\cal L}_K$ is obtained by a path integral calculation using spin coherent 
states \cite{FRADKIN}. Finally, we integrate out the field $\vec \ell$ which 
appears quadratically in $\cal L$. The resultant 
Lagrangian density is found to have an $SO(3)_L \times SO(2)_R$
symmetry and can be parametrized as
\begin{equation}
{\cal L} ~=~ {1\over 2c} ~{\rm tr}(\partial_t {\underline R}^T
\partial_t {\underline R} \ P_0) ~-~ {c\over 2} ~{\rm
tr}(\partial_x {\underline R}^T \partial_x {\underline R}\ P_1)~,
\end{equation}
where $c= J_1 S a (1+4 J_2/J_1)\sqrt{1-J_1^2/16 J_2^2}$, and
$P_0$ and $P_1$ are diagonal matrices with entries given by
\begin{eqnarray}
P_0 ~&=&~ {\rm diag} ~~\left({1\over 2g_2^2}, ~{1\over 2g_2^2}, ~{{1\over
g_1^2} - {1\over 2 g_2^2}}\right)\nonumber\\
{\rm and} \quad P_1 ~&=&~ {\rm diag} ~~\left({1\over 2g_4^2}, ~{1\over
2g_4^2}, {}~{{1\over g_3^2} - {1\over 2g_4^2}}\right) ~.
\end{eqnarray}
The couplings $g_i$ are found to be
\begin{eqnarray}
g_2^2 &=& g_4^2 ~=~ {1\over S} \sqrt{{4J_2+J_1 \over
4J_2-J_1}},\nonumber\\
g_3^2 &=& 2g_2^2, \nonumber\\
{\rm and} \quad g_1^2 &=& g_2^2 ~[~1+(1-J_1 / 2 J_2)^2 ~]~.
\label{eeight}
\end{eqnarray}
Perturbatively, there are three modes, one gapless mode with the
velocity $c g_2/g_4$ and two gapless modes with the velocity $c
g_1/g_3$. Note that the theory is not Lorentz invariant because
$g_1 g_4 \ne g_2 g_3$. However, the theory is symmetric
under $SO(3)_L \times SO(2)_R$ where the $SO(3)_L$ rotations mix
the rows of the matrix $\underline R$ and the $SO(2)_R$ rotations
mix the first two columns. (To have the full $SO(3)_L \times
SO(3)_R$ symmetry, we need $g_1 = g_2$ and $g_3 = g_4$, $i.e.$,
both $P_0$ and $P_1$ proportional to the identity matrix.) The
$SO(3)_L ~$ is the manifestation in the continuum theory of the
spin symmetry of the original lattice model.  The $SO(2)_R ~$
arises in the field theory because the ground state is planar,
and the two out-of-plane modes are identical and can mix under
an $SO(2)$ rotation. The Lagrangian is also symmetric under the
discrete symmetry parity which transforms ${\underline R}{}(x)
\rightarrow {\underline R}{}(-x) P$ with $P$ being the diagonal
matrix (-1,1,-1). An important point to note is that there is
no topological term present here (unlike the NLSM in the Neel phase) and 
hence, no apparent distinction between integer and half-integer spins. There 
is, however, a distinction due to solitons, as we will show later.

At distances of the order of the lattice spacing $a$, the values
of the couplings are given in Eq. (\ref{eeight}). At larger distance scales
$l$, the effective couplings $g_i(l)$ evolve according to the
$\beta$-functions $\beta(g_i) = d g_i/dy$ where $y = {\rm ln}
(l/a)$. We have computed the one-loop $\beta$-functions using
the background field formalism \cite{SEN}. (Note that since the
theory is not Lorentz-invariant, geometric methods cannot be used to obtain 
the $\beta$-functions \cite{AZARIA}.) The $\beta$-functions are given by
\begin{eqnarray}
\beta(g_1) ~&=&~ {g_1^3 \over {8 \pi}} \left[~{g_1^2 g_3
g_4\over g_2^2} {2\over g_1 g_4 + g_2 g_3} ~+~ 2 g_1 g_3
\ ({1\over g_1^2} - {1\over g_2^2}) ~\right],\nonumber\\
\beta(g_2) ~&=&~ {g_2^3 \over {8 \pi}} \left[~g_1^3 g_3 ({2\over
g_1^2} - {1\over g_2^2})^2 ~+~ 4 g_1 g_3 ~({1\over g_2^2} -
{1\over g_1^2}) ~\right],\nonumber\\
\beta(g_3) ~&=&~ {g_3^3 \over {8 \pi}} \left[~{g_3^2 g_1
g_2\over g_4^2} {2\over g_1 g_4 + g_2 g_3} ~+~ 2 g_1 g_3
\ ({1\over g_3^2} - {1\over g_4^2})~\right],\nonumber\\ {\rm
and} \quad 
\beta(g_4) ~&=&~ {g_4^3 \over {8 \pi}} \left[~g_3^3 g_1 ({2\over
g_3^2} - {1\over g_4^2})^2 ~+~ 4 g_1 g_3 ~({1\over g_4^2} -
{1\over g_3^2})~\right].
\end{eqnarray}
We numerically investigate the flow of these couplings
using the initial values $g_i(a)$ given in Eq.(\ref{eeight}). We find that
the couplings flow such that $g_1/g_2$ and $g_3/g_4$ approach
$1$, $i.e.$, the theory flows towards $SO(3)_L \times SO(3)_R$
and Lorentz invariance. Finally, at some length scale $\zeta$,
the couplings blow up indicating that the system has become
disordered. At one-loop, $\zeta$ depends on $J_2/J_1$ but $S$
can be scaled out. In Fig. 2, we show the numerical results for
$\ln (\zeta /a)$ versus $J_2 / J_1 ~$ for $4 J_2 / J_1 > 1$.
Note that as $4 J_2/J_1 \rightarrow 1$ from either side (the
Neel phase for integer spin or the spiral phase for any spin),
ln($\zeta/a) \rightarrow 0$, $i.e.$, the correlation length goes
through a minimum. Since $4 J_2/J_1 = 1$ separates the Neel and
spiral phases, we may call it a disorder point. (For general
$\delta$, we have a disorder line $4 J_2/J_1 + \delta^2 = 1$ and
the correlation length is minimum on the disorder line
separating the two gapped phases.)

The spiral phase is therefore disordered for any spin $S$ with a
length scale $\zeta$. Since the theory flows to the principal
chiral model with $SO(3)_L \times SO(3)_R$ invariance at long
distances, we can read off its spectrum from the exact solution
given in Ref. \cite{ANALYTIC}. The low energy spectrum consists
of a massive doublet that transforms according to the spin-$1/2$
representation of $SU(2)$. It would be interesting to verify
this by numerical studies of the model. DMRG studies
\cite{DMRG,CHITRA} of spin-$1/2$ and spin-$1$ chains
have not seen these elementary excitations so far. It is likely
that these excitations are created in pairs and a naive
computation of the energy gap would only give the mass of a
pair.  To see them as individual excitations, it would be
necessary to compute the wave function of an excited state and
explicitly compute the local spin density as was done in Ref.
\cite{WHITE} to study a one magnon state in the Neel phase.

\vskip .5 true cm
\leftline{\bf C. Solitons In The Spiral Phase}
\vskip .5 true cm

Since the field theory is based on an $SO(3)$-valued field
${\underline R}\ (x,t)$ and $\pi_1(SO(3)) = Z_2$, it allows
$Z_2$ solitons.  The classical field configurations come in two
distinct classes with soliton number equal to zero or one. If
${\underline R}_0{} (x,t)$ is a zero soliton configuration, then
a one soliton configuration is obtained as
\begin{equation}
{\underline R}_1{} (x,t) = \left(
         {\begin{array}{ccc}
         \cos\theta (x) & \sin\theta (x) & 0\\
         -\sin\theta (x) & \cos\theta (x) & 0\\
         0 & 0 & 1
         \end{array}}
         \right) {\underline R}_0{} (x,t) ~,
\end{equation}
where $\theta(x)$ goes from $0$ to $2\pi$ as $x$ goes from
$-\infty$ to $+\infty$. (For convenience, we choose $\theta(x) =
2\pi - \theta(-x)$, $i.e.$, the twist is parity symmetric about
the origin.) In terms of spins, this corresponds to
progressively rotating the spins so that the spins at the right
end of the chain are rotated by $2\pi$ with respect to spins at
the left end. Since the derivative $\partial_x \theta$ can be
made vanishingly small, the difference in the energies of the
configurations ${\underline R}_0{} (x,t)$ and ${\underline
R}_1{} (x,t)$ can be made arbitrarily small, and one might
expect to see a double degeneracy in the spectrum.

However, this classical continuum argument needs to be examined
carefully in the context of a quantum lattice model. Firstly, do
${\underline R}_0 \  (x,t)$ and ${\underline R}_1{} (x,t)$
actually correspond to orthogonal quantum states? For the spin
model, if the region of rotation is spread out over an odd
number of sites, $i.e.$, if the rotation operator is 
\beq
U = {\rm
exp} ({i\pi\over 2m+1} \sum_{n=-m}^m (2n+2m+1) S_n^z),
\eeq
then
${\underline R}_0{} (x,t)$ and ${\underline R}_1{} (x,t)$ have
opposite parities because under parity, 
\beq
S_i^z \rightarrow - S_{-i}^z \quad {\rm and} \quad 
U \rightarrow U {\rm exp}(i2\pi \sum_{n=-m}^m S_n^z).
\eeq
Since the sum contains an odd number of spins,
the term multiplying $U$ is $-1$ for half-integer spin and $1$
for integer spin. Thus for half-integer spin, ${\underline
R}_0{} (x,t)$ and ${\underline R}_1{} (x,t)$ are orthogonal and
the argument for double degeneracy of the spectrum is valid.
This is just a restatement of the Lieb-Schultz-Mattis theorem
\cite{LSM}. For integer spin, ${\underline R}_0 \ (x,t)$ and
${\underline R}_1{} (x,t)$ have the same parity, and no
conclusion can be drawn regarding the degeneracy of the spectrum.

An alternative argument leading to a similar conclusion can be
made following Haldane \cite{FDMH}. We consider a tunneling
process between a zero soliton configuration ${\underline R}_0{}
(x,t)$ and a one soliton configuration ${\underline R}_1{}
(x,t)$. (We choose coplanar configurations for convenience).
Such a tunneling process is not allowed in the continuum theory
(which is why the solitons are topologically stable) because the
configurations have to be smooth at all space-time points. But
in the lattice theory, discontinuities at the level of the
lattice spacing are allowed.  In terms of spins, this tunneling
can be brought about by turning each spin ${\bf S}_i^{(0)}$ in
configuration ${\underline R}_0{} (x,t)$ to the spin ${\bf
S}_i^{(1)}$ in configuration ${\underline R}_1{} (x,t)$ by
either a clockwise or an anticlockwise rotation. Assuming that
the magnitude of the amplitude for the tunneling is the same (as
we will show below), the contribution of the two paths either
add or cancel depending on whether the spin is integral or
half-integral. This is easily seen through a Berry phase \cite{FRADKIN} 
calculation. The difference in the Berry phase of the two paths from ${\bf
S}_i^{(0)}$ to ${\bf S}_i^{(1)}$ is $2 \pi S$. Since the soliton
involves an odd number of spins, the total Berry phase difference is $0$ mod 
$2\pi$ if $S$ is an integer and $\pi$ mod $2\pi$ if $S$ is half-integer.

Now we have to check that the magnitudes of the amplitudes for
tunneling are the same in both the cases. To see this, consider
the pair of spins ${\bf S}_{i}^{(0)}$ and ${\bf S}_{-i}^{(0)}$
which need to be rotated to ${\bf S}_i^{(1)}$ and ${\bf
S}_{-i}^{(1)}$. Since $\theta(x) = 2 \pi -\theta (-x)$, the
magnitude of the amplitude for the clockwise rotation of ${\bf
S}_{i}^{(0)}$ to ${\bf S}_i^{(1)}$ is matched by the magnitude of
the amplitude for the anticlockwise rotation of ${\bf
S}_{-i}^{(0)}$ to ${\bf S}_{-i}^{(1)}$. Hence, for the pair of
spins taken together, the magnitude of the amplitude for
tunneling is the same for the clockwise and anticlockwise rotations.

Thus, tunneling between soliton sectors is possible for integer $S$ (thereby 
breaking the classical degeneracy and leading to a unique quantum ground 
state), but not for half-integer $S$ (due to cancellations between pairs of 
paths). This agrees with the earlier Lieb-Schultz-Mattis argument.

Although the NLSM model for the spiral phase was explicitly
derived only for $\delta =0$, we expect the same qualitative
features to persist when $\delta \ne 0$, because the spin wave
analysis shows that the classical ground state continues to be
coplanar and there continue to be three zero modes (two with
identical velocities and the third with a higher velocity.
Hence we expect similar RG flows and a similar spectrum.
However, the argument for the double degeneracy of the ground
state for half-integer spins depends on parity being a good
quantum number. When $\delta \ne 0$, parity no longer commutes
with the Hamiltonian and the argument breaks down. This is in
agreement with the DMRG studies \cite{CHITRA} (for periodic
chains) which show a unique ground state, both for integer and
half-integer spins, for $\delta \ne 0$. For open chains, the
ground state is sometimes degenerate due to end degrees of
freedom. To incorporate such effects, one would have to study
NLSM theories on open chains which is beyond the scope of this work.

\vskip .5 true cm
\leftline{\bf D. Colinear Phase}
\vskip .5 true cm

Finally, we examine small fluctuations in the colinear phase.
The naive expectation is that the field theory would be an
$O(3)$ NLSM, analogous to the Neel phase, since the classical
ground state is colinear. We can show this explicitly for
$\delta = 1$ which is called the Heisenberg ladder
\cite{LADDER}. The field theory in this limit can be derived using the 
classical periodicity under translation by four lattice sites, similar to the 
derivation given in Se. IIIA  for the Neel phase. 
For a set of four neighbouring spins, we define 
\bea
\quad {{\vec \phi}}~ (x-a) = {{{\bf S}_{4i} - {\bf S}_{4i+1}}\over
2S}, ~ \quad & {{\vec \ell}}~ (x-a)& = {{{\bf S}_{4i} + {\bf
S}_{4i+1}}\over 2a}~, \nonumber \\
{\rm and} \quad  {{\vec \phi}}~ (x+a)  =  {{\bf S}_{4i+3} - {\bf
S}_{4i+2} \over 2S} ~, &{{\vec \ell}}~ (x+a)& = {{\bf S}_{4i+3} +
{\bf S}_{4i+2}\over 2a}~
\label{phi2}
\eea
where $x=(4i+3/2)a$ is the mid-point of the set of four spins.
We then write the Hamiltonian in terms of the fields ${\vec \phi}$
and ${\vec \ell}$, and Taylor expand to second order in
space-time derivatives to obtain the Lagrangian (\ref{lag1})
{\it without} a topological term. We now find
$c=4aS\sqrt{J_2(J_2+J_1)}$ and $g^2 = {1\over S} ~\sqrt{(J_2 +
J_1 )/ J_2}$. The absence of the topological term means that
there is no difference between integer and half-integer spins
and a gap exists in both cases.  In fact, the NLSM predicts a
gap for any finite inter-chain coupling, however small. This is
in agreement with numerical work on coupled spin chains \cite{LADDER}.

\vskip .5 true cm
\leftline{\bf IV. DISCUSSION}
\vskip .5 true cm

In conclusion, we emphasize that this is the first systematic
field theoretic treatment of the general $J_1-J_2-\delta$ model on a chain.  
Although all experimental spin chain systems known to date,
like NENP and $Sr_2 Cu O_3 ~$, are in the Neel phase \cite{HGAPEXPT,AMI}, 
it would be interesting to find an experimental system with sufficient 
frustration and dimerization to probe the spiral and colinear phases. These 
phases could also be studied using numerical
techniques like DMRG. The field theoretic treatment of the
spiral phase leads to the interesting possibility that the low
energy excitations of integer spin models may be massive
spin-$1/2$ objects. This again is a possibility which could be
looked for experimentally or verified by numerical simulations.

The field theories for general $\delta$ in both the spiral and 
colinear phases are still open questions. Although the results are 
qualitatively expected to be similar to the $\delta=0$ case in the spiral
phase and the $\delta=1$ case in the colinear case, quantitative features
such as the dependence of the gap on the coupling strengths will require the
explicit form of the field theory.

\vskip 1 true cm

\newpage

\noindent {\bf Figure Captions}
\vskip 1 true cm

\noindent {1.} Classical phase diagram of the $J_1 -J_2 -\delta$ spin chain.

\noindent {2.} Plot of $\ln (\zeta/a)/S$ versus $J_2/J_1$ for
$\delta =0$. For $4 J_2/J_1<1$, $\ln (\zeta/a)$ is given by the
one-loop RG of the $O(3)$ NLSM for integer spin. For $4 J_2/J_1>1$, $\ln 
(\zeta/a)$ is given by the one-loop RG of the $SO(3)_L \times SO(2)_R$ NLSM.


\begin{thebibliography}{99}
\bibitem{FN1} We use the word `phase' for convenience to denote the
position of the peak in the spin-spin correlation function $S(q)$. There is 
actually no phase transition in the spin chain even at zero temperature.
\end{thebibliography}
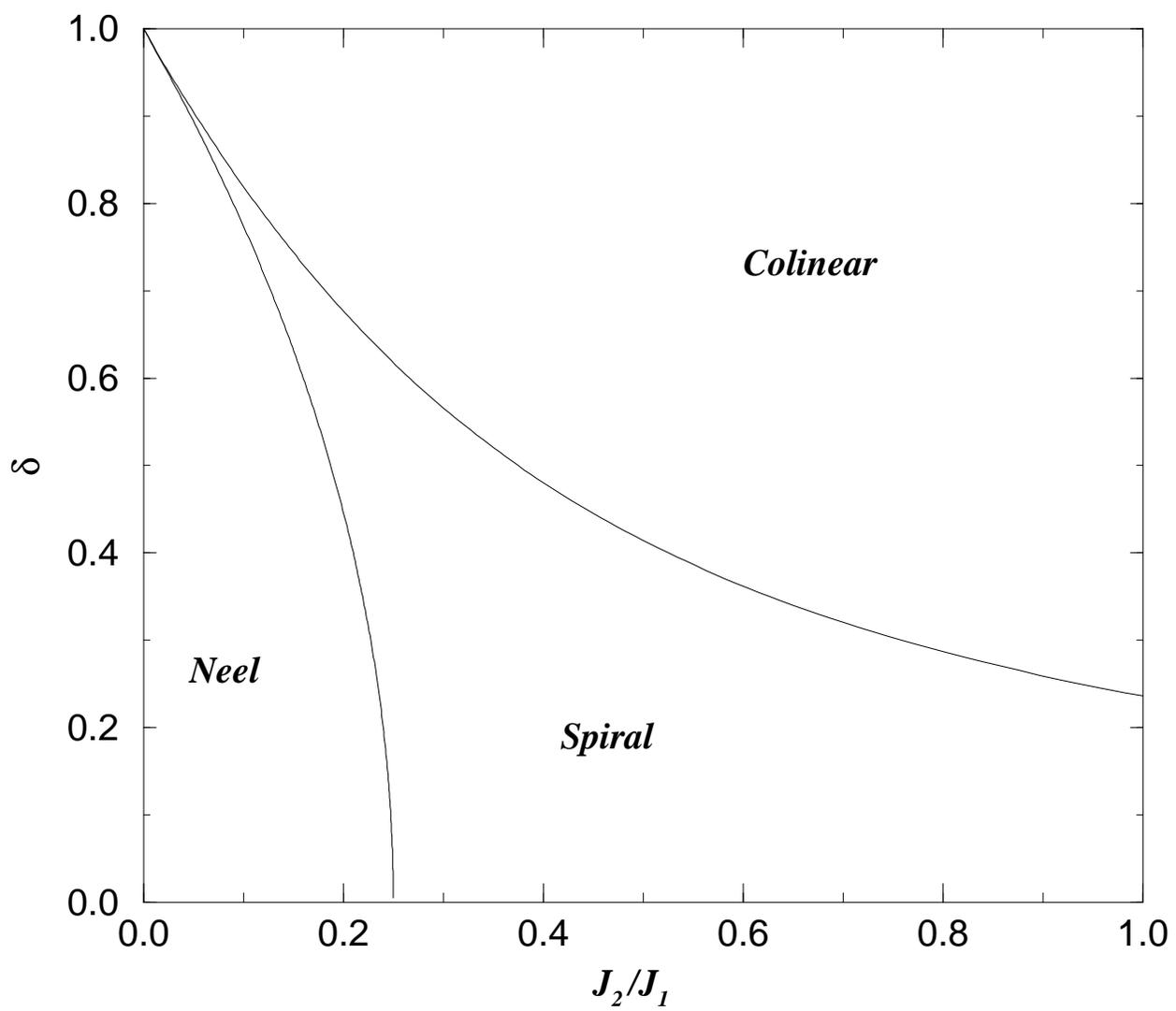
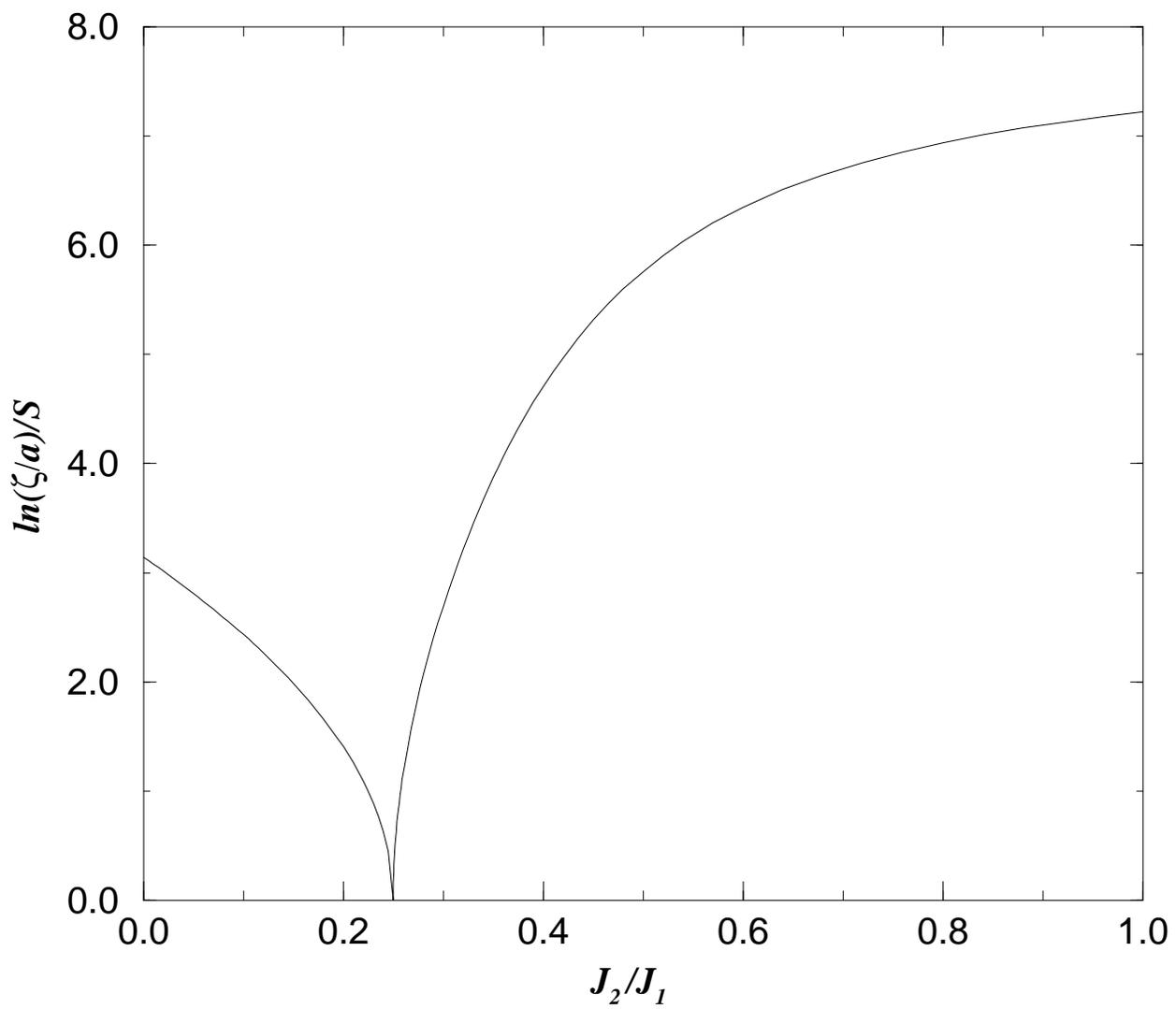
\end{document}